\documentclass[preprint,showpacs,preprintnumbers,amssymb,superscriptaddress,aps,prd,nofootinbib,11pt]{revtex4-1}

\selectfont 

\usepackage{graphicx}       
\usepackage{dcolumn}        
\usepackage{bm}             
\usepackage{psfrag}
\usepackage{tensor}
\usepackage[usenames]{color}
\definecolor{navyblue}{rgb}{0.0, 0.0, 0.5}
\usepackage[linktocpage,colorlinks=true,allcolors=navyblue]{hyperref}
\usepackage{amsmath}

\newcommand{\omtil}{\widetilde{\omega}}

\newcommand{\beq}{\begin{equation}}
\newcommand{\eeq}{\end{equation}}

\newcommand{\nn}{\nonumber}

\newcommand{\mbar}{\overline{m}}

\begin{document}

\preprint{}

 \title{Instability of the Proca field on Kerr spacetime}

\author{Sam R. Dolan}
 \email{s.dolan@sheffield.ac.uk}
\affiliation{Consortium for Fundamental Physics,
School of Mathematics and Statistics,
University of Sheffield, Hicks Building, Hounsfield Road, Sheffield S3 7RH, United Kingdom}

\date{\today}

\begin{abstract}
A massive vector boson field in the vicinity of a rotating black hole is known to suffer an instability, due to the exponential amplification of (co-rotating, low-frequency) bound states by black hole superradiance. Here we calculate the bound state spectrum by exploiting the separation of variables recently achieved by Frolov, Krtou\v{s}, Kubiz\v{n}\'ak and Santos (FKKS) for the Proca field on Kerr-(A)dS-NUT spacetimes of arbitrary dimension. Restricting to the 4D Kerr case, we first establish the relationship between the FKKS and Teukolsky variables in the massless case; obtain exact results for the angular eigenvalues in the marginally-bound case; and present a spectral method for solving the angular equation in the general case. We then demonstrate that all three physical polarizations can be recovered from the FKKS ansatz, resolving an open question. We present numerical results for the instability growth rate for a selection of modes of all three polarizations, and discuss physical implications.
\end{abstract}

\maketitle

%
%

\section{Introduction}

In a Penrose process \cite{Penrose:1971uk} a black hole can lose mass, angular momentum and/or charge, and yet still increase its horizon area, in a manner consistent with the second law of black hole mechanics \cite{Bardeen:1973gs}. An example of a Penrose process is black hole superradiance \cite{Brito:2015oca}, in which a bosonic field becomes amplified through the extraction of mass and angular momentum from a black hole. In 1972, Press and Teukolsky \cite{Press:1972zz} considered a scenario whereby a `mirror'  reflects superradiance back onto the black hole of mass $M$, triggering an instability in which the bosonic field's amplitude grows exponentially with time. A mirror is not necessary, however \cite{Damour:1976kh}. A bosonic field with a rest mass $\mu$ has a spectrum of (quasi-)bound states that are effectively trapped in the vicinity of black hole, and these states can be exponentially amplified by superradiance \cite{Detweiler:1980uk, Zouros:1979iw}. 

The superradiant instability is highly sensitive to the ratio of the gravitational radius of the black hole to the Compton wavelength of the field. For an efficient process, one requires $r_g / \lambda_C \sim O(1)$. For a rotating black hole of mass $M$, the instability could be triggered if there exists in nature an ultra-light field with a mass $\mu \lesssim 7 \times 10^{-12} \text{eV} \times \left(10M_\odot / M \right)$. This constraint is chiefly due to the fact that superradiance is a low-frequency phenomenon, associated with the angular frequency of the event horizon itself,
$
\Omega_H \approx  \frac{a}{r_+}  \left(\frac{10M_\odot}{M} \right) \times 10 \, \text{kHz} ,
$
where $a \equiv J / M$ is the spin rate of the black hole ($0 \le a < M$) and $r_+ \equiv M + \sqrt{M^2-a^2}$ is the radius of the event horizon. The first law of black hole mechanics \cite{Bardeen:1973gs} implies that 
$
d\mathcal{A} = \left(1 - \Omega_H dJ/dM \right) 8 \pi \kappa^{-1} d M,
$
where $\mathcal{A}$, $\Omega_H$ and $\kappa$ are the area, angular frequency and surface gravity of the black hole's horizon, respectively, and $dM$ and $dJ$ are changes in its mass and angular momentum (and henceforth $G=c=\hbar=1$). A mode of a bosonic field with frequency $\omega$ and azimuthal number $m$ is associated with a change $dJ/dM = m / \omega$. Then, by the second law ($d\mathcal{A} \ge 0$), the black hole will lose mass-energy ($dM \le 0$) into any field mode that satisfies the superradiant condition $\omega \omtil < 0$, where $\omtil \equiv \omega - m \Omega_H$.

The archetype for an ultra-light boson is the (hypothetical) axion, a pseudoscalar introduced to solve the strong CP problem of QCD \cite{Peccei:1977hh}. String-theory-inspired theories can generate axion-like particles, with masses that are not linked to the axion decay constant. Compactifications could lead to a generic landscape of ultralight axions, known as the ``string axiverse'' \cite{Arvanitaki:2009fg}, populating all mass scales possibly down to the present Hubble scale. Massive hidden $U(1)$ vector fields are also a generic feature in BSM scenarios, particularly in string theory compactifications, where such fields arise from e.g.~broken non-Abelian orbifolds in heterotic compactifications, and D-brane configurations and bulk Ramond-Ramond fields in type II string theories \cite{Goodsell:2009xc,Jaeckel:2010ni}. Ultra-light bosonic fields with masses $\ll eV$ are considered as plausible dark-matter candidates \cite{Duffy:2009:axions, Baker:2013:quest}; see for example the recent hypothesis of a scalar with mass $\sim 10^{-22} \text{eV}$ \cite{Hui:2016ltb}. 

Ultralight bosons in the dark sector -- if extant -- should trigger superradiant instabilities with potentially observable consequences for astrophysical black holes \cite{Arvanitaki:2010sy, Kodama:2011zc, Yoshino:2013ofa, Brito:2017wnc, Brito:2017zvb}, such as (i) gaps in the black hole mass-spin plane, to be revealed by black hole surveys \cite{Arvanitaki:2010sy, Pani:2012vp, Pani:2012bp, Baryakhtar:2017ngi, Cardoso:2018tly}; (ii) gravitational wave `sirens', contributing to the stochastic background or resolvable in their own right \cite{Arvanitaki:2009fg, Brito:2017wnc}, and (iii) significant transfers of mass-energy from the black hole into a surrounding bosonic `cloud' \cite{Hod:2012px,Herdeiro:2014goa,Herdeiro:2015waa,Herdeiro:2016tmi}, of up to $\sim 9\%$ \cite{East:2017ovw}. The prospect that black holes could act as astrophysical particle detectors is an intriguing one \cite{Brito:2014wla}. Astronomical datasets are already being used to put upper bounds on the populations and masses of ultra-light bosons in the dark sector \cite{Cardoso:2018tly, Stott:2018opm}. Primordial black holes, if they exist, will also create signatures via superradiant instabilities: for example, a $M \sim 10^{24} \text{kg}$ black hole with an axion of mass $\mu \sim 10^{-5} \text{eV}/c^2$ could generate millisecond-bursts in the GHz radio-frequency range \cite{Rosa:2018igv}. 

Whether or not it is realised in nature, the superradiant instability is of theoretical interest. 
A field of mass $\mu$ surrounding a black hole admits a discrete spectrum of `quasi-bound states'. These are modes with harmonic time dependence $\exp(-i \omega t)$ that are regular on the future (outer) horizon $r_+$, and which fall away exponentially far from the black hole. As this system is `open' at the horizon, the frequencies are complex: $\omega = \omega_R + i \omega_I$. For Schwarzschild black holes, all modes decay through the horizon and thus $\omega_I < 0$. Conversely, for Kerr black holes, any modes which satisfy the superradiance condition  have a positive imaginary component, $\omega_I > 0$, as they grow exponentially with time, with an e-folding time of $\omega_I^{-1}$.  

In the small-$M\mu$ regime, the bound state spectrum is approximately hydrogenic, 
\beq
\frac{\omega_R}{\mu} \approx 1 - \frac{M\mu^2}{2n^2} + O(M \mu)^4 ,
\eeq 
with fine and hyperfine structure corrections at $O(M\mu)^4$ and $O((am/M) (M\mu)^5)$, respectively \cite{Ternov:1980st, Detweiler:1980uk, Lasenby:2002mc, Dolan:2015eua, Baumann:2018vus}. 
The growth rate of the dominant mode is a strong power of $M\mu$, viz.~\cite{Detweiler:1980uk, Rosa:2011my, Pani:2012vp, Pani:2012bp, Endlich:2016jgc, Baryakhtar:2017ngi, Cardoso:2018tly}, 
\beq
M \omega_I \approx  (2 r_+ \gamma_{\ell m S \hat{n}}) (m \Omega_H - \omega_R) (M\mu)^{4 |m| + 5+ 2S} , \quad \quad (M\mu \ll 1). \label{eq:omI}
\eeq
Here $\gamma_{\ell m S \hat{n}}$ is a coefficient which depends on the mode numbers, and $S$ represents the polarization state of the field (with $S=0$ for a scalar field). For the Proca field, $S$ takes the values $\{-1,0,1\}$ for the three polarizations, describing the relationship between orbital and spin angular momentum \cite{Rosa:2011my,Pani:2012vp}.

For $M\mu \gg 1$, the low-$m$ modes lie outside the superradiant regime, but growth can still occur in high-$m$ modes, though its rate is exponentially-suppressed with $M\mu$ \cite{Zouros:1979iw}. Numerical calculations \cite{Furuhashi:2004jk,Cardoso:2005vk,Dolan:2007mj} have found that the growth is most rapid in the dipole ($m=1, \ell =1$) for $\omega_R$ (and thus $\mu$) close to the cut-off frequency of $\Omega_H < 1/(2M)$. 

Over the last decade there have been steps towards calculating the spectrum of the Proca field on Kerr spacetime \cite{Rosa:2011my, Witek:2012tr, Pani:2012bp,Pani:2012vp, Endlich:2016jgc, East:2017mrj, East:2017ovw, Baryakhtar:2017ngi} (see also \cite{Konoplya:2005hr, Konoplya:2006gq}), leading up to first precise numerical results reported in 2018 \cite{Cardoso:2018tly}. The Proca field instability has a much faster maximum rate than the scalar field instability for two reasons: superradiance is enhanced with field spin, and the $S = -1$ dipole mode is bound more tightly to the black hole than the scalar-field dipole mode. Eq.~(\ref{eq:omI}) shows that the $S=-1$ mode has the smallest index, and thus it is expected to grow parametrically faster than a scalar field, and the other polarizations of the Proca field. 

Finding bound states of the Proca field has been a technical challenge, due to the apparent inseparability of the governing equations: after assuming harmonic dependence in $t$ and $\phi$, one is still left with a coupled set of PDEs in $r$ and $\theta$ for the components of the vector potential. In 2012, Pani \emph{et al.} \cite{Pani:2012bp,Pani:2012vp} addressed the slow-rotation ($a \ll M$) regime using series expansion methods. In 2017, Baryakhtar \emph{et al.} \cite{Baryakhtar:2017ngi} used separable approximations in the near-horizon and far-field regimes, and a matching argument; and East \& Pretorius \cite{East:2017mrj, East:2017ovw} studied the fullly non-linear system with a numerical relativity code, taking into account the back-reaction of the Proca field on the spacetime geometry. In 2018, Cardoso \emph{et al.}~\cite{Cardoso:2018tly} developed a numerical approach to solving the coupled PDEs directly, providing accurate numerical data for the $m=1$, $S=-1$ mode for the first time. 

Recently, Frolov, Krtou\v{s}, Kubiz\v{n}\'ak \& Santos (FKKS) \cite{Frolov:2018ezx, Krtous:2018bvk} have shown something remarkable: the equations governing the Proca field on the Kerr-(A)dS-NUT spacetimes of arbitary dimension are \emph{separable}, once a certain ansatz is employed, inspired by the work of Lunin \cite{Lunin:2017drx}. The problem of finding bound state modes reduces to that of solving a pair of ordinary differential equations. FKKS have computed the growth rate for the `even-parity' modes $S = -1$ and $S = +1$ for $a = 0.998M$ \cite{Frolov:2018ezx}. 

Here we show that the `odd-parity' modes $S = 0$ also emerge from the FKKS ansatz, with the subtlety that the separation constant $\nu$ \footnote{Here $\nu$ is used in place of $\mu$ in FKKS \cite{Frolov:2018ezx}, as we use $\mu$ to denote the mass of the Proca field} diverges in the limit $a \rightarrow 0$, though in such a way as to leave a regular radial equation. Thus we present results for all three physical polarizations for the first time.

The paper is organised as follows. In Sec.~\ref{sec:setup} we review the symmetries of the Kerr spacetime (\ref{sec:symmetries}) and the approach of FKKS to separability (Sec.~\ref{sec:fkks}). We obtain separable expressions for the Maxwell scalars $\phi_0$ and $\phi_1$ (Sec.~\ref{sec:maxwell}) and examine the link to the Teukolsky formalism in the massless limit $\mu = 0$ (Sec.~\ref{sec:teukolsky}). In Sec.~\ref{sec:method} we outline the approach to finding the quasi-bound state spectrum. First, we introduce a spectral-decomposition method (Sec.~\ref{sec:spectral}) for finding the eigenvalues $\nu$ of the angular equation. This approach leads to exact results (Sec.~\ref{sec:exact}) for the angular solutions in the special case $\omega^2 = \mu^2$. The numerical method for finding bound states is outlined in Sec.~\ref{sec:radial}. A selection of numerical results are presented and interpreted in Sec.~\ref{sec:results}. We conclude with a short discussion in Sec.~\ref{sec:discussion}.


\section{Proca field on Kerr spacetime\label{sec:setup}}

\subsection{Kerr spacetime and its symmetries\label{sec:symmetries}}
The Kerr spacetime, in the exterior of the outer horizon, is most commonly expressed in terms of Boyer-Lindquist coordinates $\{t,r,\theta,\phi\}$. An alternative choice \cite{Frolov:2017kze} is $\{ \tau = t - a \phi, r, y = a \cos \theta, \psi = \phi / a\}$ and the line element 
\beq
ds^2 = g_{a b} dx^a dx^b = -\frac{\Delta_r}{\Sigma} \left(d \tau + y^2 d \psi \right)^2 + \frac{\Delta_y}{\Sigma} (d \tau - r^2 d\psi)^2 + \frac{\Sigma}{\Delta_r} dr^2 + \frac{\Sigma}{\Delta_y} dy^2 , 
\eeq
where $\Sigma \equiv \sqrt{-g} = r^2 + y^2$ and $\Delta_r \equiv r^2 - 2Mr + a^2$, $\Delta_y \equiv a^2 - y^2 = a^2 \sin^2\theta$. 

The metric $g_{ab}$ admits a pair of Killing vectors $\xi_{(\tau)}^a = \partial^a_\tau = \partial^a_t$ and $\xi_{(\psi)}^a = \partial^a_\psi = a \partial_\phi + a^2 \partial_\tau$. The spacetime also admits a closed conformal Killing-Yano tensor $h_{ab} = h_{[ab]}$, known as the \emph{principal tensor}, with the key property $\nabla_c h_{ab} = g_{ca} \xi^{(\tau)}_b - g_{cb} \xi^{(\tau)}_a$ and thus $\xi^a_{(\tau)} = \frac{1}{3} \nabla_b h^{ba}$, where $\nabla_a$ denotes the covariant derivative. The principal tensor can be written in terms of a potential, $h_{ab} = (db)_{ab} = 2 \nabla_{[a} b_{b]}$, where
\beq
b_a dx^a = -\frac{1}{2} \left( (r^2 - y^2) d\tau + r^2 y^2 d\psi \right)
\eeq
i.e.,
\beq
h = y dy \wedge (d \tau - r^2 d \psi) - r dr \wedge (d\tau + y^2 d\psi). 
\eeq
The Hodge dual of $h$ is $f \equiv {}^\star h$ (i.e.~$f_{ab} = \frac{1}{2} \varepsilon_{abcd} h^{cd}$),
\beq
f = r dy \wedge (d\tau - r^2 d\psi) + y dr \wedge (d\tau + y^2 d\psi) .
\eeq
The Killing-Yano tensor $f$ has the properties $f_{ab;c} = f_{[ab;c]}$, and thus $\nabla^a f_{ab} = 0$ and $\Box f_{ab} = \tensor{R}{_a ^c _b ^d} f_{cd}$. 

The Killing tensor is $K_{ab} \equiv f_{ac} \tensor{f}{_b ^c}$. The conformal Killing tensor is $Q_{ab} \equiv h_{ac} \tensor{h}{_b ^c}$. The Killing tensor satisfies $K_{(ab;c)} = 0$, and the conformal Killing tensor satisfies $Q_{(a b ; c)} = g_{(ab} q_{c)}$, with $q_c \equiv h_{cd} \xi_{(\tau)}^d$. The Killing tensor and the two Killing vectors are related by $\tensor{K}{^a _b} \xi_{(\tau)}^b = - \xi_{(\psi)}^a$. Explicit expressions for these tensors are given in Appendix \ref{appendix:a}.

\subsection{The Proca equation: separation of variables\label{sec:fkks}} 
The Proca field, describing a massive vector boson \cite{Proca:1900nv}, is governed by the field equation
\beq
\nabla_b F^{ab} + \mu^2 A^a = 0, \quad \quad F_{ab} \equiv \nabla_a A_b - \nabla_b A_a, \label{eq:Proca}
\eeq
where $\nabla_a$ denotes the covariant derivative, and $A_a$ is the vector potential. By acting on (\ref{eq:Proca}) with $\nabla_a$, it follows that $\nabla_a A^a = 0$, and thus the vector potential is necessarily in Lorenz gauge. As this removes gauge freedom, it follows that the Proca particle has three physical polarizations, rather than the two polarizations of a massless vector boson such as the photon. On a Ricci-flat spacetime (such as Kerr), the Proca equation is equivalent to $\Box A^a - \mu^2 A^a = 0$, subject to the Lorenz-gauge constraint.

FKKS \cite{Frolov:2018ezx} employ for the vector field $A^a$ the ansatz \cite{Lunin:2017drx} 
\beq
A^a = B^{ab} \nabla_b Z , \label{eq:ansatz}
\eeq 
where $Z$ is a scalar function which can be written in the multiplicative separated form
\beq
Z = R(r) S(\theta) e^{-i \omega t} e^{i m \phi} ,
\eeq 
and $B^{ab}$ is a tensor field satisfying \cite{Krtous:2018bvk}
\beq
B^{ab} (g_{bc} + i \nu h_{bc}) = \delta_c^a ,  \label{eq:Bdef}
\eeq
with $\nu$ a constant to be determined. 


FKKS have shown that, on the Kerr-(A)dS-NUT family of spacetimes, with this ansatz, Eq.~(\ref{eq:Proca}) is satisfied if $R(r)$ and $S(\theta)$ obey a pair of second-order ordinary differential equations \cite{Krtous:2018bvk, Frolov:2018ezx}. In the Kerr case, these equations are
\begin{subequations}
\begin{eqnarray}
q_r \frac{d}{dr} \left[ \frac{\Delta}{q_r} \frac{dR}{dr} \right] + \left[ \frac{K_r^2}{\Delta} + \frac{2-q_r}{q_r} \frac{\sigma}{\nu} - \frac{q_r \mu^2}{\nu^2} \right] R(r) &=& 0 ,  \\
 \frac{q_\theta}{\sin \theta} \frac{d}{d\theta} \left[ \frac{\sin \theta}{q_\theta} \frac{dS}{d\theta} \right] - \left[ \frac{K_\theta^2}{\sin^2 \theta} + \frac{2-q_\theta}{q_\theta} \frac{\sigma}{\nu} - \frac{q_\theta \mu^2}{\nu^2} \right] S(\theta) &=& 0 , 
\end{eqnarray} 
\label{eq:FKKS}
\end{subequations}
where 
\begin{align}
K_r &= (a^2+r^2)\omega - am, & K_\theta &= m - a\omega \sin^2 \theta, & \Delta &= \Delta_r = r^2 - 2Mr + a^2 = (r-r_+)(r-r_-),  \nn \\
q_r &= 1+\nu^2 r^2, & q_\theta &= 1 - \nu^2 a^2 \cos^2 \theta, & \sigma &= \omega + a\nu^2 (m-a\omega). \label{eq:sigma}
\end{align}

\subsection{Maxwell scalars\label{sec:maxwell}}
By solving Eq.~(\ref{eq:Bdef}), one obtains an explicit expression for the tensor $B^{ab}$ in Eq.~(\ref{eq:ansatz}) given by 
\beq
B^{ab} = \frac{1}{\Sigma} \left[  \frac{1}{(1 + \nu^2 r^2)} \begin{pmatrix} -r^4 \Delta_r^{-1} & i \nu r^3 & 0 & -r^2 \Delta_r^{-1} \\
-i \nu r^3 & \Delta_r & 0 & -i\nu r \\ 0 & 0 & 0 & 0 \\  -r^2 \Delta_r^{-1} & i \nu r & 0 & -\Delta_r^{-1}\end{pmatrix} + \frac{1}{(1-\nu^2 y^2)} \begin{pmatrix} y^4 \Delta_y^{-1} & 0 & i \nu y^3 & -y^2 \Delta_y^{-1} \\ 0 & 0 & 0 & 0 \\ -i\nu y^3 & 0 & \Delta_y & i \nu y \\ -y^2 \Delta_y^{-1} & 0 & -i\nu y & \Delta_y^{-1} \end{pmatrix} \right] ,
\label{eq:Bab}
\eeq
with respect to the coordinates $\{\tau , r , y , \psi \}$. The Lorenz condition $\nabla_a A^a = 0$ is found to separate with separation constant $\kappa \equiv \mu^2 / \nu^2 - \omega / \nu$, yielding Eqs.~(\ref{eq:FKKS}). 

The Maxwell scalars are defined by $\phi_0 \equiv F_{ab} l^a m^b$, $\phi_1 \equiv \frac{1}{2} F_{ab} \left(l^a n^b + \mbar^a m^b \right)$ and $\phi_2 = F_{ab} \mbar^a n^b$, where 
\beq
l^a = \Delta_r^{-1} \left[ r^2, \Delta_r, 0, 1 \right], \quad
n^a = (2\Sigma)^{-1} \left[ r^2, -\Delta_r, 0, 1 \right] , \quad
m^a = \frac{1}{\sqrt{\Delta_y}} \frac{1}{r + i y} \left[-i y^2, 0, -\Delta_y, i \right] , \label{eq:kinnersley}
\eeq
are null vectors of the Kinnersley tetrad, and $\mbar^a$ is the complex conjugate of $m^a$. With equations (\ref{eq:Bab}) and (\ref{eq:kinnersley}), one obtains
\begin{subequations}
\begin{eqnarray}
\phi_0 &=& \phantom{-} \left(\frac{i \nu}{\sqrt{2}} \right) \left( \frac{\mathcal{D}_0 R}{1+ i \nu r} \right)  \left( \frac{\mathcal{L}_0^\dagger S}{1+ a \nu \cos \theta} \right) , \\
2(r-ia \cos \theta)^2 \phi_2 &=& - \left(\frac{i \nu}{\sqrt{2}} \right) \left( \frac{\Delta \mathcal{D}_0^\dagger R}{1- i \nu r} \right)  \left( \frac{\mathcal{L}_0 S}{1- a \nu \cos \theta} \right) .
\end{eqnarray}
\label{eq:maxwell}
\end{subequations}
where
\begin{subequations}
\begin{align}
\mathcal{D}_n &= \partial_r - i \Delta_r^{-1} K_r + n \Delta_r^{-1} \partial_r \Delta_r, &
\mathcal{L}_n &= \partial_\theta + m \csc \theta - a \omega \sin \theta + n \cot \theta, \\
\mathcal{D}_n^\dagger &= \partial_r + i \Delta_r^{-1} K_r + n \Delta_r^{-1} \partial_r \Delta_r, &
\mathcal{L}_n^\dagger &= \partial_\theta - m \csc \theta + a \omega \sin \theta + n \cot \theta .
\end{align}
\end{subequations}
Thus the Maxwell scalars $\phi_0$ and $\phi_2$ can be written in separable form. The expression for $\phi_1$ is somewhat longer and is omitted here.

The equations (\ref{eq:FKKS}) may be rearranged into the form
\begin{subequations} \label{eq:R1S1}
\begin{eqnarray}
\frac{d}{dr} \left[ \Delta \frac{dR}{dr} \right] + \left[ \frac{K_r^2}{\Delta} - \Lambda + 2a\omega m - a^2 \omega^2 - \mu^2 r^2 \right] R &=& \frac{2 r \nu^2}{q_r} \left[ \Delta \frac{d}{dr} + r \frac{\sigma}{\nu}  \right] R  ,  \label{eq:R1} \\
 \frac{1}{\sin \theta} \frac{d}{d\theta} \left[ \sin \theta \frac{dS}{d\theta} \right] + \left[ \Lambda - \frac{m^2}{\sin^2 \theta} + a^2 \gamma^2 \cos^2 \theta \right] S &=& \frac{2 a^2 \nu^2 \cos \theta}{q_\theta} \left[\sin \theta \frac{d}{d\theta} + \frac{\sigma}{\nu} \cos \theta \right] S , \label{eq:S1} 
\end{eqnarray}
\end{subequations}
where $\Delta = \Delta_r$, 
\beq
\Lambda(\nu) \equiv \mu^2 / \nu^2 - \sigma / \nu + 2 a \omega m - a^2 \omega^2 , \label{eq:Lambda}
\eeq 
and $\gamma^2 \equiv \omega^2 - \mu^2$. Here $\Lambda$ is a function of $\nu$, rather than a free parameter in its own right. The motivation for the rearrangement (\ref{eq:R1S1}) is that the left-hand sides are now equivalent in form to the equations governing the massive scalar field \cite{Brill:1972xj}.

\subsection{Massless limit: Teukolsky equations\label{sec:teukolsky}}
In the massless limit $\mu \rightarrow 0$, one may establish a connection to the Teukolsky equations for $s = \pm 1$ \cite{Teukolsky:1972my, Teukolsky:1973ha, Press:1973zz}. The massless case has been investigated in detail by Lunin \cite{Lunin:2017drx}. 

Teukolsky's approach uses a separable ansatz for the Maxwell scalars, viz.,
\begin{subequations}
\begin{align}
\phi_0 &= R_{+1}(r) S_{+1}(\theta) e^{-i \omega t + i m \phi}, \\
2 (r-i a \cos \theta)^2 \phi_2 &= R_{-1}(r) S_{-1}(\theta) e^{-i \omega t + i m \phi} ,
\end{align}
\label{eq:teuk}
\end{subequations}
which leads to ordinary differential equations for the functions $R_{\pm1}$ and $S_{\pm1}$,
\begin{subequations}
\begin{align}
\left(\Delta \mathcal{D}_0^\dagger \mathcal{D}_0 - 2 i \omega r \right) R_{-1} &= \lambda R_{-1} , &
(\mathcal{L}_0 \mathcal{L}_1^\dagger + 2 a \omega \cos \theta) S_{-1} = -\lambda S_{-1} , \\
\left(\Delta \mathcal{D}_0 \mathcal{D}_0^\dagger + 2 i \omega r \right) \Delta R_{+1} &= \lambda \Delta R_{+1} , &
(\mathcal{L}_0^\dagger \mathcal{L}_1 - 2 a \omega \cos \theta) S_{+1} = -\lambda S_{+1} ,
\end{align}
\label{eq:teukolsky-eqns}
\end{subequations}
where $\lambda$ is the separation constant for $s=-1$ \cite{Chandrasekhar:1998}.

By comparing Eq.~(\ref{eq:teuk}) with Eq.~(\ref{eq:maxwell}), we read off the Teukolsky radial functions $R_{\pm1}$ as
\beq
R_{+1} = \frac{i \nu}{\sqrt{2} C_+} \frac{\mathcal{D}_0 R}{(1 + i \nu r)} , \quad \quad
R_{-1} = -\frac{i \nu}{\sqrt{2} C_-} \frac{\Delta \mathcal{D}^\dagger_0 R}{(1 - i \nu r)}
\label{eq:Rpm}
\eeq 
and the spin-weighted spheroidal harmonics $S_{\pm1}$ as
\beq
S_{+1} = \frac{C_{+} \mathcal{L}_0^\dagger S}{(1 + a\nu \cos \theta)} ,  \quad \quad 
S_{-1} = \frac{C_{-} \mathcal{L}_0 S}{(1 - a\nu \cos \theta)} ,
\label{eq:Spm}
\eeq 
for some choice of normalization constants $C_{\pm}$.   
Using the FKKS equations (\ref{eq:FKKS}) with $\mu^2 = 0$, one may show that these functions do indeed satisfy the Teukolsky equations (\ref{eq:teukolsky-eqns}) if and only if we make the identification
\beq
\lambda = -\frac{\omega}{\nu} + (m - a \omega) a \nu ,
\eeq
that is, $\lambda = (\sigma - 2 \omega) / \nu$. 
Inverting this relationship gives
\beq
\nu = \frac{\lambda \pm \mathcal{B}}{2 a (m-a\omega)} = \frac{-2 \omega}{\lambda \mp \mathcal{B}},
\label{eq:nu-massless}
\eeq
where $\mathcal{B} \equiv \sqrt{\lambda^2 + 4 a m \omega - 4 a^2 \omega^2}$ is the Teukolsky-Starobinsky constant \cite{Press:1973zz, Starobinsky:1973aij, Starobinskii:1973amp}. A further useful relationship is $\sigma / \nu =  \omega/\nu + (m-a\omega) a\nu = \pm \mathcal{B}$. 

Note that one eigenvalue $\lambda$ yields two eigenvalues $\nu$ here; in expressions below the $\pm$ sign corresponds to the choice made here. The transformation $\nu \rightarrow -\omega/ (m - a\omega) (a\nu)^{-1}$ leaves $\lambda$ unchanged, and changes the sign of $\sigma / \nu$. 


Via similar steps, and with the choice  $C_+ / C_- = \pm 1$, one may (re-)establish the Teukolsky-Starobinsky identities
\begin{subequations}
\begin{align}
\Delta \mathcal{D}_0 \mathcal{D}_0 R_{-1} &= \mathcal{B} \Delta R_{+1} , &
\mathcal{L}_0^\dagger  \mathcal{L}_1^\dagger  S_{-1} &= \mathcal{B} S_{+1} , \\
\Delta \mathcal{D}_0^\dagger \mathcal{D}_0^\dagger \Delta R_{+1} &= \mathcal{B} R_{-1} , &
\mathcal{L}_0  \mathcal{L}_1  S_{+1} &= \mathcal{B} S_{-1} .
\end{align}
\end{subequations}
The inverse of Eqs.~(\ref{eq:Rpm}) and (\ref{eq:Spm}), giving the FKKS functions $R(r)$ and $S(\theta)$ in terms of the Teukolsky functions, are
\begin{subequations}
\begin{align}
\pm \mathcal{B} R  &= \phantom{-\;\;} (1+i\nu r) \mathcal{D}_0 \hat{R}_{-1} - i \nu \hat{R}_{-1} , \\
&= - \left[ (1-i\nu r)  \mathcal{D}_0^\dagger \Delta \hat{R}_{+1} + i \nu \Delta \hat{R}_{+1} \right] ,
\end{align}
\end{subequations}
and 
\begin{subequations}
\begin{align}
\pm \mathcal{B} S &= (1 + a \nu \cos \theta) \mathcal{L}_1^\dagger \hat{S}_{-1} + a \nu \sin \theta \hat{S}_{-1} , \\
&= (1 - a \nu \cos \theta) \mathcal{L}_1 \hat{S}_{+1} - a \nu \sin \theta \hat{S}_{+1} , 
\end{align}
\end{subequations}
where $\hat{R}_{+1} \equiv C_+ R_{+1}$, $\hat{R}_{-1} \equiv C_- R_{-1}$, $\hat{S}_{+1} \equiv S_{+1}/C_+$ and $\hat{S}_{-1} \equiv S_{-1}/C_-$. 

Lunin \cite{Lunin:2017drx} identified an electric and magnetic polarization $A_a^{(el)}$ and $A_a^{(mag)}$, with eigenvalues $\nu^{(el)}$ and $\nu^{(mag)}$, respectively, derived from separable functions $Z^{(el)}(\nu^{(el)})$ and $Z^{(mag)}(\nu^{(mag)})$. Applying the duality transformation $\nu^{(mag)} = -1/a\nu^{(el)}$, Lunin showed that $Z^{(el)}$ and $Z^{(mag)}$ satisfy the same differential equations. Going one step further, one can show that $A_a^{(mag)} = a \nu^{(el)} A_a^{(el)} - a\nu^{(el)} \partial_a Z$; thus the electric and magnetic polarizations are equivalent up to a gauge transformation, in the massless case.

We note that FKKS use a different ansatz for $B^{ab}$ than that used by Lunin for the electric polarization. Here we take $A^a = B^{ab} \partial_b Z$, with $B^{ab}$ given in Eq.~(\ref{eq:Bab}). Lunin uses $A^a_{(el)} = B^{\prime ab} \partial_b Z_{(el)}$, where $B^{\prime ab} = (i \nu)^{-1} ( g^{ab} - B^{ab} )$. However, for $\mu = 0$, this leads to vector potentials $A^a$ and $A^a_{(el)}$ which differ only by a multiplicative constant and a gauge term. 

\section{Method\label{sec:method}}
Below we outline our approach to solving the radial and angular equations to find the spectrum of quasi-bound states of the Proca field on Kerr spacetime.

\subsection{Solving the angular equation\label{sec:spectral}}
The separation constant $\nu$ is found by solving the angular equation (\ref{eq:S1}) subject to imposing regularity on $A^a$ across the domain $0 \le \theta \le \pi$, including at the poles ($\theta = 0$, $\theta = \pi$). 

We shall allow $\nu$ is take any value. At first glance, real values of $\nu$ such that $|a \nu| > 1$ would appear to cause a divergence in $A^a$, due to the factor of $1 - \nu^2 y^2$ appearing in the denominator of the second term of $B^{ab}$ in Eq.~(\ref{eq:Bab}). However, in the limit $y \rightarrow \nu$ (i.e.~$q_\theta \rightarrow 0$), the vanishing of the term in square brackets on the right-hand side of (\ref{eq:S1}) ensures that $B^{ab} \nabla_b Z$ remains regular. Therefore, $A^a$ is regular away from the poles if $S(\theta)$ is also regular. 

Here we employ a spectral decomposition method, similar to that used for solving the spin-weighted spheroidal harmonic equation that arises in Teukolsky's equations \cite{Hughes:1999bq, Cook:2014cta}. We expand the function $S(\theta)$ in the basis of spherical harmonics $Y_j^{m}(\theta, \phi) =Y_j^{m}(\theta) e^{im \phi}$, viz.,
\beq
S(\theta) = \sum_{k' = 0}^\infty b_{k'} Y_{\ell'}^{m}(\theta), \quad \quad \ell' \equiv |m| + 2 k' + \eta , \label{eq:Sansatz}
\eeq
where $\eta = 0$ or $1$. The angular equation does not couple harmonics of opposite parity, and so an eigensolution $S(\theta)$ takes a definite parity, and thus is expanded in only either odd or even $\ell$-modes.

First, multiplying Eq.~(\ref{eq:S1}) by $q_\theta$ and rearranging leads to 
\beq
(1-\nu^2 a^2 c^2) \left[\frac{d^2}{d\theta^2} + \frac{c}{s} \frac{d}{d\theta} - \frac{m^2}{s^2} + \Lambda \right] S +  \left[ \left( \gamma^2 - 2 \sigma \nu \right) a^2 c^2 -  \gamma^2 \nu^2 a^4 c^4 - 2 \nu^2 a^2 s c \frac{d}{d\theta} \right] S = 0 . \label{eq:S2}
\eeq
where $s \equiv \sin \theta$ and $c \equiv \cos \theta$. 
Inserting Eq.~(\ref{eq:Sansatz}) into Eq.~(\ref{eq:S2}) and integrating against $2 \pi \int_0^\pi \sin(\theta) \overline{Y}^{m}_{\ell} (\theta)$, where $\ell = |m| + 2 k + \eta$, leads to
\beq
\sum_{k'=0}^\infty \mathcal{M}_{kk'} b_{k'} = 0
\label{eq:M}
\eeq
where
\beq
M_{kk'} \equiv
\left[ \Lambda - \ell' (\ell' + 1) \right] \delta_{\ell \ell'} + \left[ - \nu^2 \Lambda + \nu^2 \ell' (\ell' + 1) - 2 \sigma \nu + \gamma^2 \right] a^2 c^{(2)}_{\ell \ell'} - 2 a^2 \nu^2 d^{(4)}_{\ell \ell'} - \gamma^2 \nu^2 a^4 c^{(4)}_{\ell \ell'} .
\eeq
with $\ell \equiv |m| + 2 k + \eta$ and $\ell' \equiv |m| + 2 k' + \eta$. 
Here
\begin{subequations}
\begin{eqnarray}
c^{(2)}_{\ell \ell'} &\equiv& \left<\ell m | \cos^2 \theta | \ell' m \right> = 
\frac{2\sqrt{\pi}}{3} \left<\ell, 0, \ell'\right> + \frac{4}{3}\sqrt{\frac{\pi}{5}} \left<\ell, 2, \ell'\right>
, \\
c^{(4)}_{\ell \ell'} &\equiv& \left<\ell m | \cos^4 \theta | \ell' m \right> = 
\frac{2\sqrt{\pi}}{5} \left<\ell, 0, \ell'\right> + \frac{8}{7} \sqrt{\frac{\pi}{5}} \left<\ell, 2, \ell'\right> + \frac{16 \sqrt{\pi}}{105} \left<\ell, 4, \ell'\right> 
, \\
d^{(2)}_{\ell \ell'} &\equiv& \left<\ell m \left| \sin \theta \cos \theta \frac{d}{d\theta} \right| \ell' m \right> \nn \\
&=& \sqrt{\frac{4 \pi}{3}} \left( \ell' \sqrt{\frac{(\ell'+1)^2 - m^2}{(2\ell'+1)(2\ell'+3)}} \left< \ell, 1, \ell' + 1\right> - (\ell'+1) \sqrt{\frac{\ell'^2 - m^2}{(2 \ell' + 1)(2\ell' - 1)}} \left<\ell, 1, \ell' - 1\right> \right)
\end{eqnarray}
\end{subequations}
where $\left<\ell m | \hat{X} | \ell' m \right> \equiv \int_\Omega (Y^{m}_\ell)^\ast \hat{X} Y^{m}_{\ell'} d \Omega$ with $d\Omega = \sin \theta d\theta d\phi$, and
\beq
\left<\ell_1, \ell_2, \ell_3 \right> \equiv (-1)^{m} \sqrt{\frac{(2\ell_1+1)(2\ell_2+1)(2\ell_3+1)}{4\pi}} \begin{pmatrix} \ell_1 & \ell_2 & \ell_3 \\ 0 & 0 & 0 \end{pmatrix} \begin{pmatrix} \ell_1 & \ell_2 & \ell_3 \\ -m & 0 & m \end{pmatrix}
\eeq
where $\begin{pmatrix} \cdot & \cdot & \cdot \\ \cdot & \cdot & \cdot \end{pmatrix}$ denote the Wigner 3-j symbols. 

The couplings $c^{(2)}_{\ell \ell'}$ and $d^{(2)}_{\ell \ell'}$ are zero for $|k-k'|>1$, and the couplings $c^{(4)}_{\ell \ell'}$ are zero for $|k-k'| > 2$. Thus, $\mathcal{M}_{kk'}$ is a band-diagonal matrix, with terms on the leading, sub-leading and sub-sub-leading diagonals, in general. In the special case of $\gamma = 0$, $\mathcal{M}_{kk'}$ is a tridiagonal matrix. 

Non-trivial solutions to Eq.~(\ref{eq:M}) arise for choices of $\nu$ such that $\text{det} \left| \mathcal{M}_{kk'} \right|
=0$. In general, one may find the roots numerically to obtain $\nu$. We now examine two special cases.

\subsubsection{Static case ($a = 0$)}

In the limit $a \rightarrow 0$, the matrix $\mathcal{M}_{kk'}$ is diagonal, and the solution satisfying the boundary conditions at the poles ($\theta = 0$, $\theta = \pi$) is given by $\Lambda(\nu) = \ell (\ell + 1)$ with $S = Y_{\ell m}(\theta)$ the spherical harmonic. In the massive case ($m \neq 0$) the solutions of Eq.~(\ref{eq:Lambda}) for $\nu$ are
\beq
\nu = \begin{cases} 
\mu^2 / \omega, & \ell = 0 , \\ 
-\frac{\omega}{\ell (\ell + 1)} \frac{1 \pm \sqrt{1 + 4 \ell (\ell + 1) \mu^2 / \omega^2}}{2} , & \ell > 0.
\end{cases}
\label{eq:static-eig}
\eeq
This gives the even-parity solutions for the Schwarzschild case: a pair of modes for $\ell >0$, and the monopole mode for $\ell = 0$. In the massless case, $\nu = - \omega / (\ell (\ell + 1))$. 
Though the odd-parity mode is apparently missing here, it can be recovered by considering the limit $a\rightarrow 0$ more carefully, as we show below. 

\subsubsection{Massless case ($\mu = 0$)}
In the massless case ($\mu = 0$), one can use the link to the Teukolsky functions established in Sec.~\ref{sec:teukolsky}. The symbol $\lambda$ corresponds to the separation constant for $s=-1$ in e.g.~Ref.~\cite{Teukolsky:1973ha,Press:1973zz}. For each $\lambda$, a pair of eigenvalues $\nu$ follow from Eq.~(\ref{eq:nu-massless}). In the static limit $a\omega = 0$, we have $\lambda = \ell (\ell + 1) = \mathcal{B}$, and one of the pair of $\nu$ is divergent. Series expansions for $\lambda$ in powers of $a \omega$ are given in e.g.~Ref.~\cite{Press:1973zz,Seidel:1988ue,Berti:2005gp}: use e.g.~Eqs.~(2.13)--(2.16) in Ref.~\cite{Berti:2005gp} with $s=-1$ and $\lambda = A - 2ma\omega + a^2 \omega^2$.

\subsubsection{Marginally-bound case: $\omega^2 = \mu^2$\label{sec:exact}}
In the case $\gamma = 0$, the matrix $\mathcal{M}_{kk'}$ is tridiagonal. We may seek special solutions with a terminating power series expansion, i.e.,
\beq
S = Y_{\ell}^{m}(\theta) + b_1 Y_{\ell+2}^{m}(\theta), \quad \quad \ell = |m| + \eta .  \label{Sansatz}
\eeq
Inserting (\ref{Sansatz}) into (\ref{eq:S2}) yields three equations, $\varepsilon_k = 0$ for $k=0,1,2$, where $\varepsilon_k \equiv \sum_{k'} \mathcal{M}_{kk'} b_{k'}$ with $b_0=1$ and $b_2 = 0$. 

For $\eta=0$ there is an exact solution for the $m = \pm \ell$ modes, where $\ell$ is any positive integer. The exact solution takes the form (\ref{Sansatz}) with $b_1=0$ and $\eta=0$. Making the choice 
\beq
\nu = \frac{ \mp \omega}{m - a \omega},  \label{eq:nu1}
\eeq
one finds that Eq.~(\ref{eq:Lambda}) yields $\Lambda = \ell (\ell + 1)$, and Eq.~(\ref{eq:sigma}) yields $\sigma / \nu = \mp m$. It follows that the right-hand side of Eq.~(\ref{eq:S1}) is the azimuthal-raising ($m>0$) or lowering ($m < 0$) operator. The raising/lowering operator annihilates $P_\ell^{\pm \ell}(\cos \theta)$, which is also solution of the left-hand side of (\ref{eq:Lambda}) as $\Lambda = \ell (\ell + 1)$. Thus, $P_\ell^{\pm \ell}(\cos \theta)$ is a valid solution to Eq.~(\ref{eq:S1}) with eigenvalue (\ref{eq:nu1}). We shall observe later that this eigenvalue corresponds to the polarization state $S = -1$.

The choice $b_1 = 0$ and $\eta = 1$ in Eq.~(\ref{Sansatz}), with  $m = \pm (\ell - 1)$, yields two non-trivial equations, $\varepsilon_0=0$  and $\varepsilon_1 = 0$. However, $\varepsilon_0$ and $\varepsilon_1$ share a common factor of
$
a \nu^2 + (a\omega \mp \ell) \nu - \omega ,
$
yielding a pair of roots, 
\beq
\nu = \frac{1}{2a} \left(\pm \ell - a \omega + \epsilon \sqrt{(\mp \ell  + a\omega)^2 + 4 a \omega} \right), \label{eq:S0eig}
\eeq
with $\epsilon^2 = 1$. The choice $\epsilon = \mp 1$ gives an eigenvalue $\nu$ which reduces to $\mp \omega / \ell$ in the limit $a \rightarrow 0$. The static eigenvalues (\ref{eq:static-eig}) also reduce to $\mp \omega / \ell$ in the limit $\gamma \rightarrow 0$, suggesting that we have identified even-parity modes here.

On the other hand, the choice $\epsilon = \pm 1$ yields eigenvalues that diverge in the static limit $a \rightarrow 0$. Yet the following limits are well-defined:
\beq
\lim_{a \rightarrow 0} (a \nu) = \pm \ell , \quad \quad \lim_{a \rightarrow 0}  \frac{\sigma}{\nu} = \ell (\ell - 1)  = -\lim_{a \rightarrow 0} \Lambda.
\eeq
Taking the $a \rightarrow 0$ limit of the radial equation (\ref{eq:R1}), multiplying by $r^2 f$, where $f(r) = 1 - 2M/r$,  and noting that $q'_r / q_r \rightarrow 2/r$ as $\nu^2 \rightarrow \infty$, leads to 
\beq
f \frac{d}{dr} \left[  f \frac{dR}{dr} \right] + \left( \omega^2 - f \left( \frac{\ell (\ell - 1)}{r^2} + m^2 \right) \right) R = 0. 
\eeq
This we recognise as the odd-parity Schwarzschild radial equation: see Eq.~(14) of Ref.~\cite{Rosa:2011my}), with $l$ in \cite{Rosa:2011my} replaced by $\ell - 1$ here, and $u_{4}(r) \leftrightarrow R(r)$. In other words, we have identified Eq.~(\ref{eq:S0eig}) with $\epsilon = \pm 1$ and specifying the eigenvalue $\nu$ corresponding to the $S=0$ mode for $m = \pm (\ell - 1)$. That is, to find the odd-parity dominant $m = 1$ mode, we take $\ell = 2$ here, but $l = 1$ in Ref.~\cite{Rosa:2011my}. The eigenvalue $\nu$ diverges in the limit $a \rightarrow 0$, but nevertheless produces a well-defined radial equation in this limit. 

The third solution, corresponding to the $S = +1$ polarization, is a solution with $\eta = 0$ but $b_1 \neq 0$ (for $a>0$). For $m=1$ this is the middle root of the cubic
\beq
a \nu^3 (1 - a\omega) - (6 - a \omega(2 - a\omega))\nu^2 + \omega \nu + \omega^2 .
\eeq

\subsubsection{General case}
In the general case, $a > 0$ and $\gamma \neq 0$, one does not have closed forms for the angular eigenvalues. In principle, one could look for a two-parameter series expansion for $\nu$ in e.g.~$a\omega$ and $a \gamma$. In this work we were content to find the eigenvalue numerically, by searching for roots of $\text{det} \left( \mathcal{M}_{kk'} \right)$ over the (complex) $\nu$ domain. When looking for bound states, the results of the previous section typically provide good starting guesses for $\nu$. This is because $\gamma^2 \equiv \omega^2 - \mu^2 \approx - M^2 \mu^4 / n^2$, and so $\gamma$ is small in the hydrogenic regime (i.e.~for small $M\mu$).

\subsection{Solving the radial equation\label{sec:radial}}
Bound states of the radial equation (\ref{eq:R1}) are defined by the following asymptotic conditions:
\beq
R(r) \sim \begin{cases} 
  e^{-i \omtil r_\ast} , & r \rightarrow r_+ , \\
  r^{(2\omega^2 - \mu^2)/Q} e^{-Q r} , & r \rightarrow \infty ,
\end{cases}
\eeq
where $Q \equiv \sqrt{\mu^2 - \omega^2}$ and $dr_\ast / dr \equiv (r^2+a^2) / \Delta$. 
We employed a direct integration method, starting near the horizon at $r = r_+ + \hat{\epsilon}$ with a typical value of $\hat{\epsilon} = 10^{-4}M$. We obtain initial conditions from a Frobenius series of the form
\beq
R(r) = x^{-i \kappa} \left(1 + c_1 x + c_2 x^2 + \ldots  \right), \quad \quad x \equiv \frac{r - r_+}{r_+ - r_-}, \quad \kappa \equiv \frac{2Mr_+ \omtil}{r_+ - r_-} ,
\eeq
where the series coefficients $c_k$ are determined from the radial equation.
Next, we integrate (\ref{eq:R1}) outwards from the near-horizon region to a suitably large radius, typically $r_\text{max} = 60 M / (M\mu)^2$. Numerical approximations for the bound state frequencies are found by seeking the local minima of $\text{log} |R(r_\text{max})|^2$ in the complex frequency domain. Starting guesses are provided by the hydrogenic approximation, $\omega / \mu \approx 1 - (M\mu)^2 / 2 n^2$, where the principal quantum number $n$ is $n = |m| + \hat{n} + S + 1$, with $\hat{n} = 0, 1, \ldots$ is the overtone number.

\section{Results\label{sec:results}}
Here we present a selection of numerical results for the bound states of the three polarizations ($S = -1,0,1$) of the Proca field on Kerr spacetime, focussing particularly on the instability in the $m=1$ modes. 

\begin{figure}
\begin{center}
 \includegraphics[width=10cm]{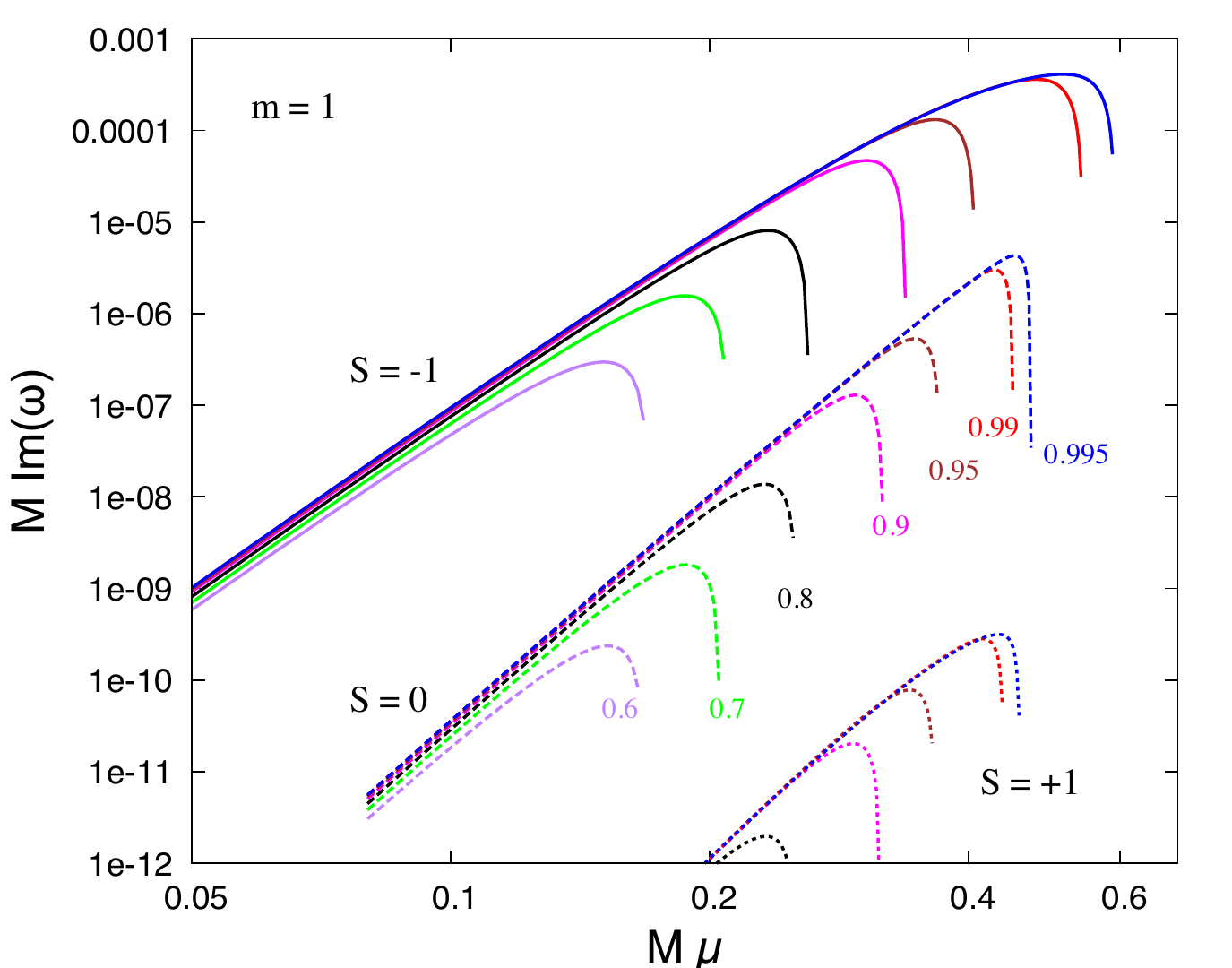}
\end{center}
\caption{
Growth rate of the fundamental ($\hat{n} = 0$) corotating $m = 1$ modes of the Proca field, for the three polarizations $S = -1$ [solid], $S = 0$ [dashed] and $S = +1$ [dotted], and for BH spins of $a/M \in \left\{0.6,0.7,0.8,0.9,0.95,0.99,0.995 \right\}$. The vertical axis shows the growth rate $\tau^{-1} = (GM/c^3) \text{Im}(\omega)$ on a logarithmic scale, and the horizontal axis shows $M \mu$.
}
\label{fig:m1S}
\end{figure}

Figure \ref{fig:m1S} shows the growth rate of the superradiant instability in the fundamental $m=1$ modes for all three polarizations, $S = -1$, $0$ and $1$. The $S=-1$ mode is dominant (fastest-growing), followed by $S = 0$, then $S = +1$. The growth rates differ greatly, with $\gtrsim 2$ orders of magnitude between $S = -1$ and $S=0$, and $\gtrsim 4$ orders of magnitude between $S = 0$ and $S = +1$. In the $M\mu \rightarrow 0$ regime, the growth rate has a power-law scaling, with an index that depends on $S$ as described by Eq.~(\ref{eq:omI}). The instability cuts off once $\omega_R$ exceeds the angular frequency of the horizon $\Omega_H$, and so $\omega_I$ changes sign. The plot illustrates how this cut-off changes with the black hole spin rate $a/M$, leading to a large difference in maximum growth rate between moderate spins (e.g.~$a/M = 0.6$) and the near-extremal case.

\begin{figure}
\begin{center}
 \includegraphics[width=10cm]{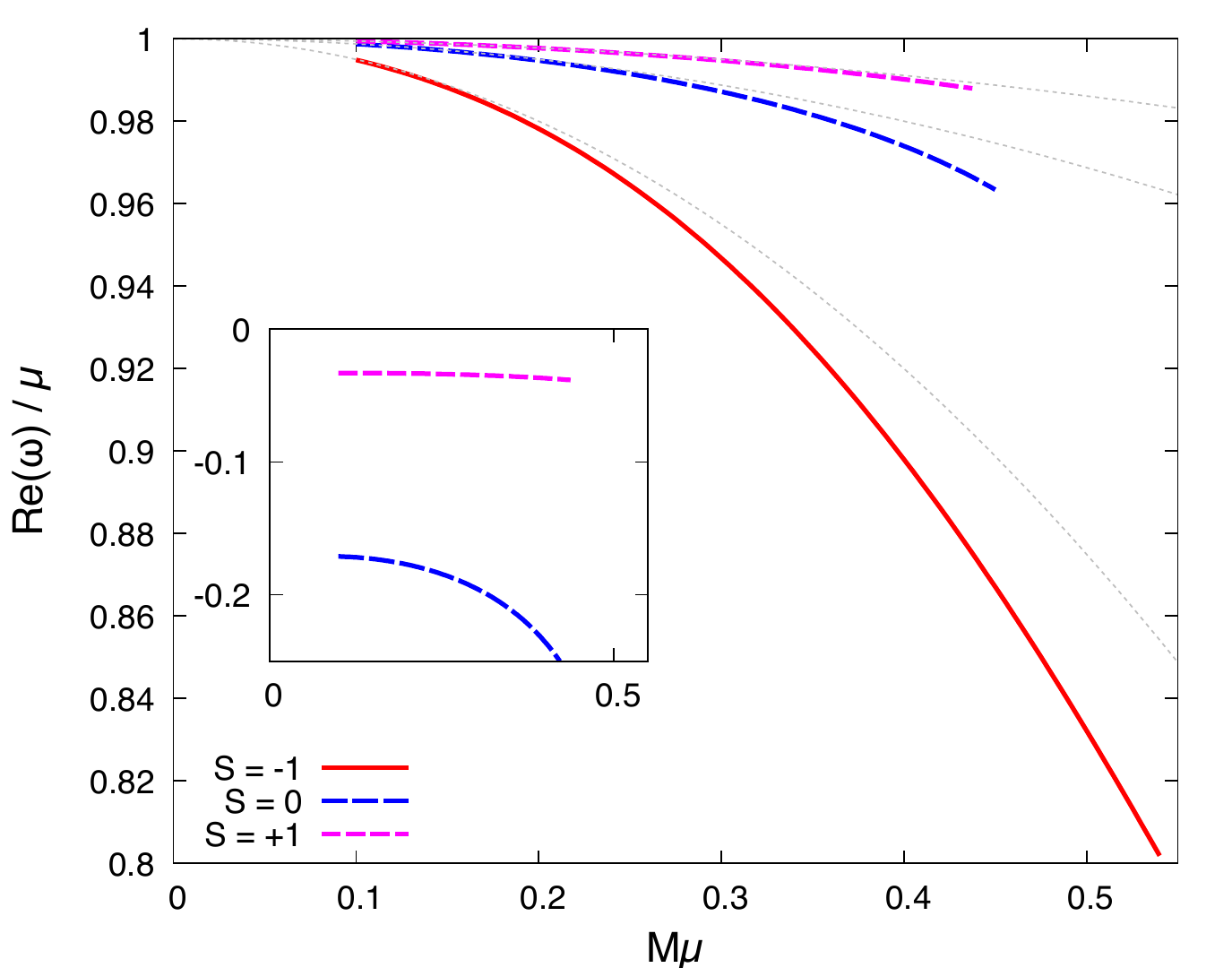}
\end{center}
\caption{
The quasi-hydrogenic spectrum of the Proca bound states. The red, blue and magenta lines show the binding energy $\text{Re}(\omega)/\mu$ for the fundamental $S=-1, 0, 1$ modes, respectively, for $a/M = 0.99$ and $m=1$. The guidelines show the hydrogenic spectrum $E_H = 1 - (M\mu)^2/(2n^2)$ for $n = 1$, $2$ and $3$. The inset shows the scaled difference, $(\text{Re}(\omega)/\mu - E_H) (M\mu)^{-4}$, that is, numerical data for the coefficient of the hyperfine structure term.
}
\label{fig:re}
\end{figure}

Figure \ref{fig:re} shows the real part of the frequency for the fundamental $m=1$ modes. In the regime $M \mu \rightarrow 0$, we observe a hydrogenic-like spectrum, with $\omega \approx \mu (1 - (M\mu^2)/2n^2)$, with a principal number $n = |m| + S + \hat{n} + 1$. For moderate $M\mu$ there is evidence of fine and hyperfine structure corrections at $O(M\mu^4)$ and $O((am/M) (M\mu)^5)$, respectively (see \cite{Baumann:2018vus} for an analysis of the scalar field case). 

\begin{figure}
\begin{center}
 \includegraphics[width=10cm]{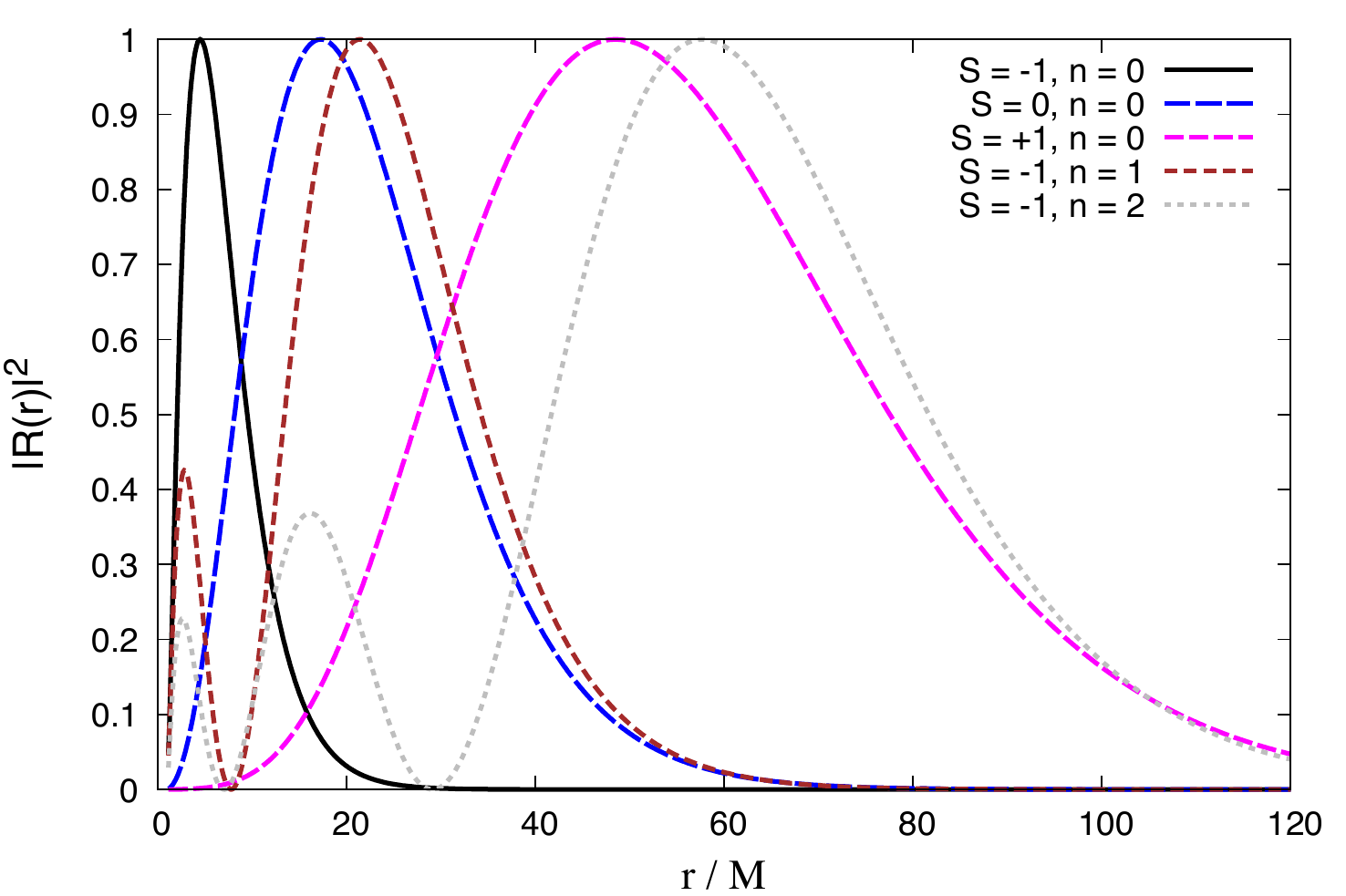}
\end{center}
\caption{
The radial profile of the bound states at $M\mu = 0.4$ and $a/M = 0.99$. Numerical parameters for these modes are given in Table \ref{tbl:params}.
}
\label{fig:wf}
\end{figure}

Figure \ref{fig:wf} shows the radial profile of the bound states for $a = 0.99$ and $M\mu = 0.4$. Numerical values for $\omega$ and $\nu$ are listed in Table \ref{tbl:params}. The fundamental modes for the three polarization have a qualitatively similar profile, with a single maximum in $|R(r)|^2$. Notably, the $S = -1$ (which is fastest-growing) is closer to the black hole, and has the largest (relative) amplitude near the horizon at $r_+$. The plot also shows the profiles of higher overtones of the $S = -1$ mode. As in the hydrogen case, the higher overtones have additional maxima \& minima. Comparing the $S = +1$ fundamental mode with the second overtone of $S = -1$, we see that, although they have a similar spatial extent, the latter has a much larger amplitude near the horizon. We therefore expect the latter to grow much more rapidly than the former; this expectation is supported by Fig.~\ref{fig:wf}.

\begin{table}
\begin{tabular}{| l | l | l | l | l | l |}
\hline
$S$ & $\hat{n}$ & $\omega_R$ & $\omega_I$ & $\nu_R$ & $\nu_I$ \\
\hline
$-1$ & $0$ & $0.35920565$ & $2.36943 \times 10^{-4}$ & $-0.60028373$ & $-3.336873 \times 10^{-4}$ \\
$0$ & $0$ &  $0.38959049$ & $2.15992 \times 10^{-6}$ & $1.84437950$ & $-9.497225 \times 10^{-7}$ \\
$+1$ & $0$ & $0.39606600$ & $2.565202 \times 10^{-10}$ & $0.21767208$ & $-3.367574 \times 10^{-11}$ \\
\hline
$-1$ & $1$ & $0.38814529$ & $5.996892 \times 10^{-5}$ & $-0.64326832$ & $-9.391193 \times 10^{-5}$ \\
$-1$ & $2$ & $0.39509028$ & $1.781747 \times 10^{-5}$ & $-0.65428679$ & $-2.863793 \times 10^{-5}$ \\
\hline
\end{tabular}
\caption{Parameters for the bound state modes shown in Fig.~\ref{fig:wf}, with $m=1$, $a/M = 0.99$ and $M\mu = 0.4$.}
\label{tbl:params}
\end{table}

\begin{figure}
\begin{center}
 \includegraphics[width=10cm]{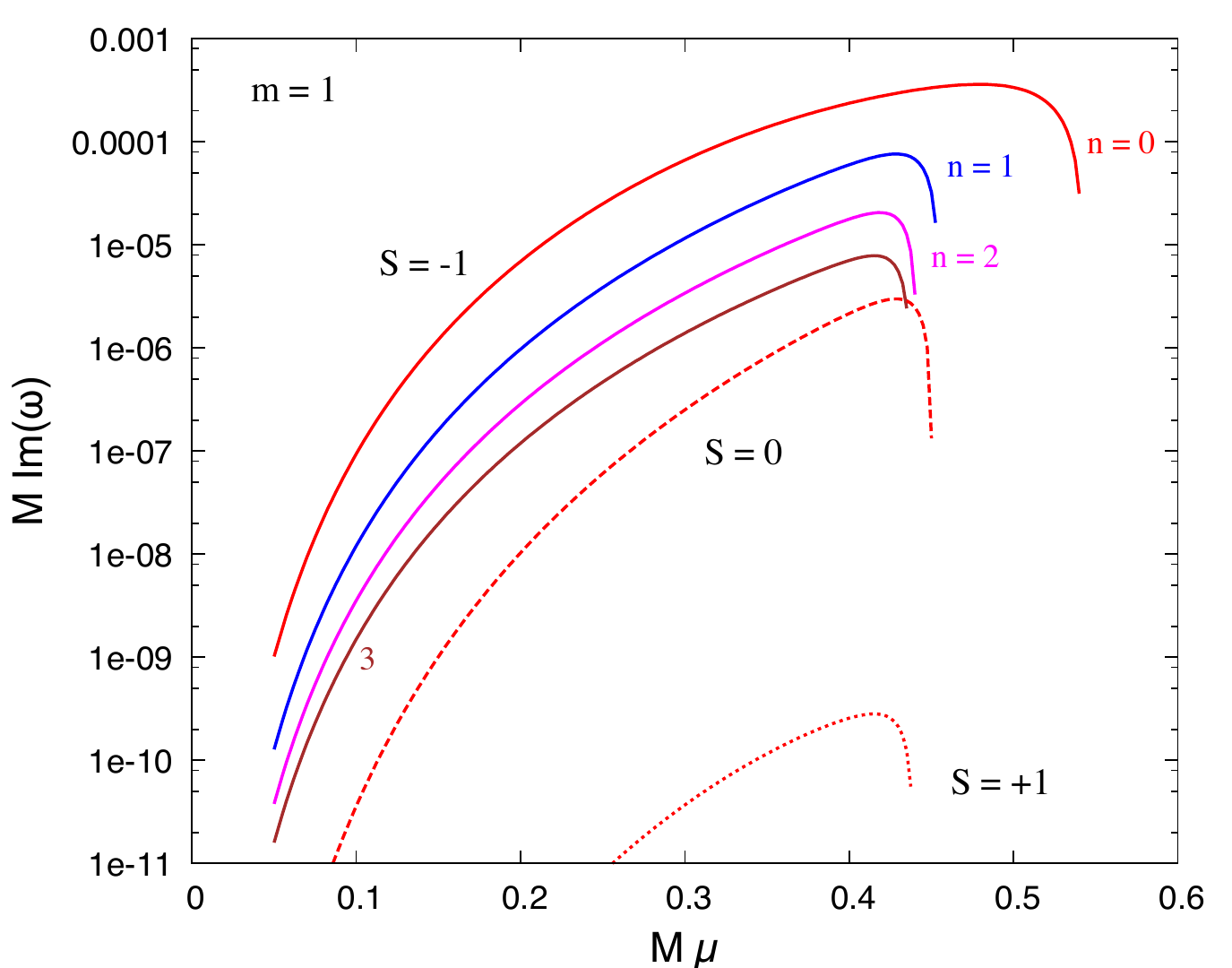}
\end{center}
\caption{
The growth rate of the higher overtones for $m = 1$ and $a = 0.99M$. The solid lines show the growth rate of the fundamental (red, $n=0$) and the higher overtones ($\hat{n} = 1,2,3$) of the $S = -1$ mode. The dashed and dotted lines show the growth rate of the fundamental modes for $S=0$ and $S=+1$, respectively.
}
\label{fig:novertone}
\end{figure}

Figure \ref{fig:novertone} shows the growth rate of the first four overtones of the $S=-1$, $m=1$ mode, and compares this against the growth rate of the $S = 0$, $m =1$ mode (see \cite{Dolan:2007mj}). It shows that several overtones of the dominant $S = -1$ mode will grow more substantially more rapidly than the fundamental mode. In essence, this is because in Eq.~(\ref{eq:omI}) the coefficient depends on the overtone $\hat{n}$ and polarization $S$, whereas the index depends only on $S$.

\begin{figure}
\begin{center}
 \includegraphics[width=10cm]{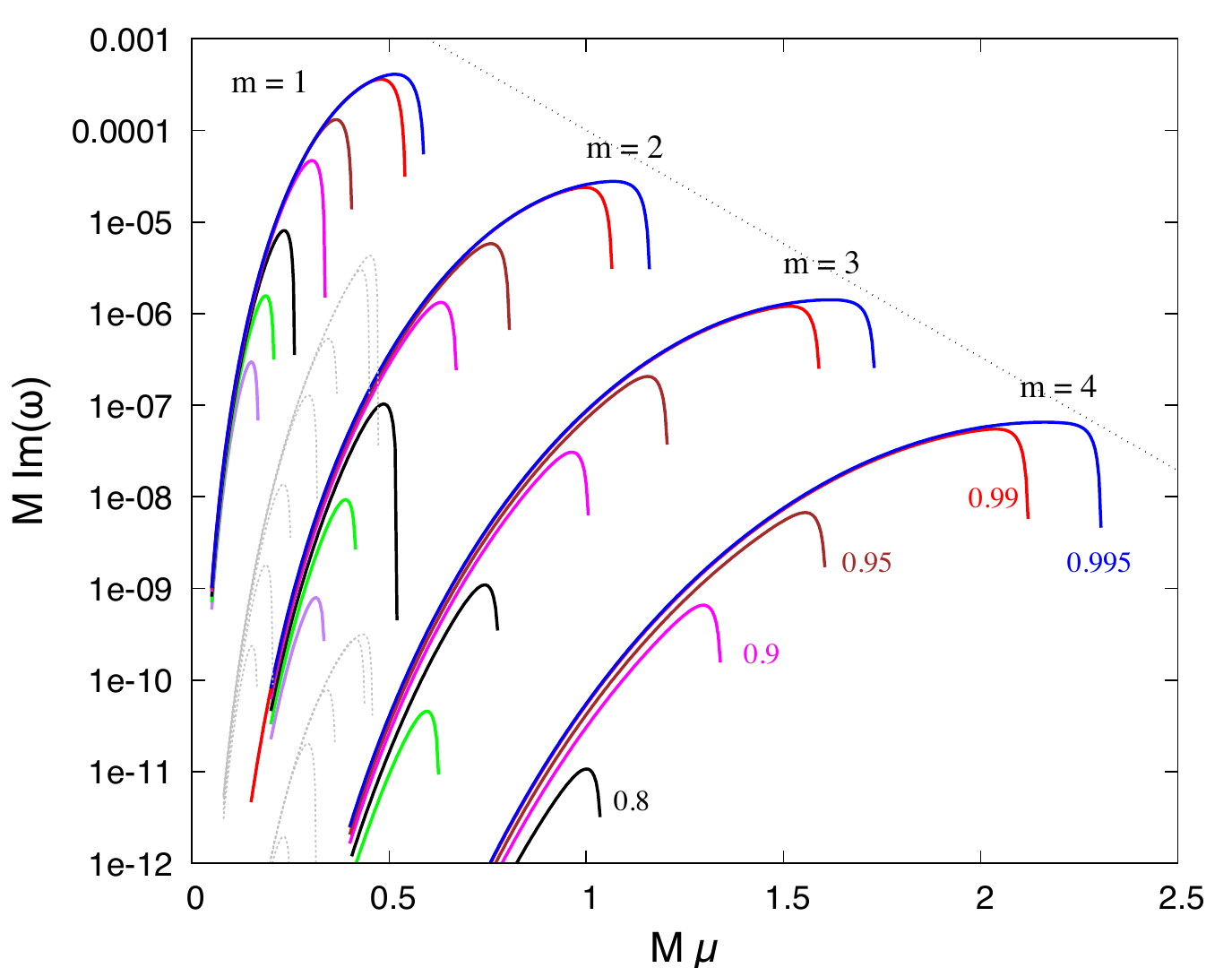}
\end{center}
\caption{
Instability growth rates for $m  = 1$, $2$, $3$ and $4$. The solid lines show the $S=-1$ modes for $a \in \{0.6,0.7,0.8,0.9,0.95,0.99,0.995\}$. The dotted lines show the $S = 0$ and $S = -1$ modes for $m=1$. 
}
\label{fig:mhigher}
\end{figure}

Figure \ref{fig:mhigher} shows the growth rate of the higher modes of the $S = -1$ polarization with azimuthal numbers $m = 2$, $3$ and $4$. The superradiant instability persists for higher modes at larger values of $M\mu$, but the rate becomes insignificant for $M\mu \gg 1$, due to the exponential fall off of $M \omega_I^{\text{max}}$ with $M\mu$ seen in Fig.~\ref{fig:mhigher}.
 
\begin{figure}
\begin{center}
 \includegraphics[width=10cm]{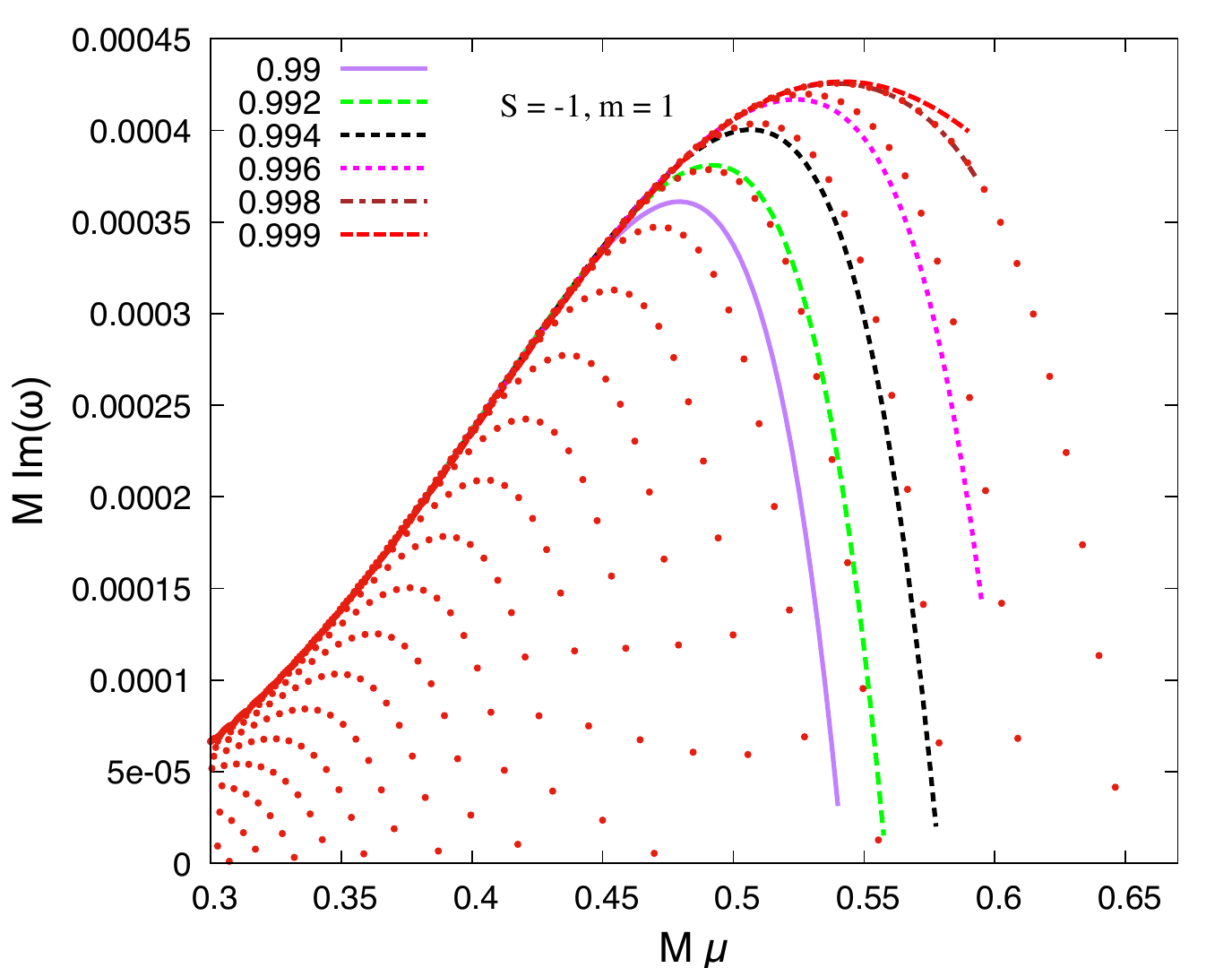}
\end{center}
\caption{
The maximum growth rate for the dominant $S = -1$, $m=1$ mode. 
The lines show new data obtained here by solving the ODEs. The points show the dataset \footnote{\url{https://arxiv.org/src/1801.01420v1/anc}} that Cardoso {\it et al.} \cite{Cardoso:2018tly} obtained by solving PDEs.
}
\label{fig:max}
\end{figure}

Figure \ref{fig:max} highlights the maximum growth rate for the dominant $S=-1$, $m=1$ mode. For $a=0.999M$, we find a maximum growth rate of $M \omega_I \approx 4.27 \times 10^{-4}$ which occurs at $M \mu \approx 0.542$. This corresponds to a minimum e-folding time of $\tau_{min} \approx 2.34\times 10^3 (GM/c^3)$. For comparison, a numerical estimate of the minimum e-folding time of the scalar field is $\tau_{min} \approx 5.88 \times 10^6 (GM/c^3)$, which occurs for the dipole mode of the scalar field at $M \mu = 0.45$ and $a = 0.997M$ \cite{Dolan:2012yt}. In other words, the Proca field instability has a maximum rate $\approx 2500$ times faster than the scalar field.

\begin{figure}
\begin{center}
 \includegraphics[width=10cm]{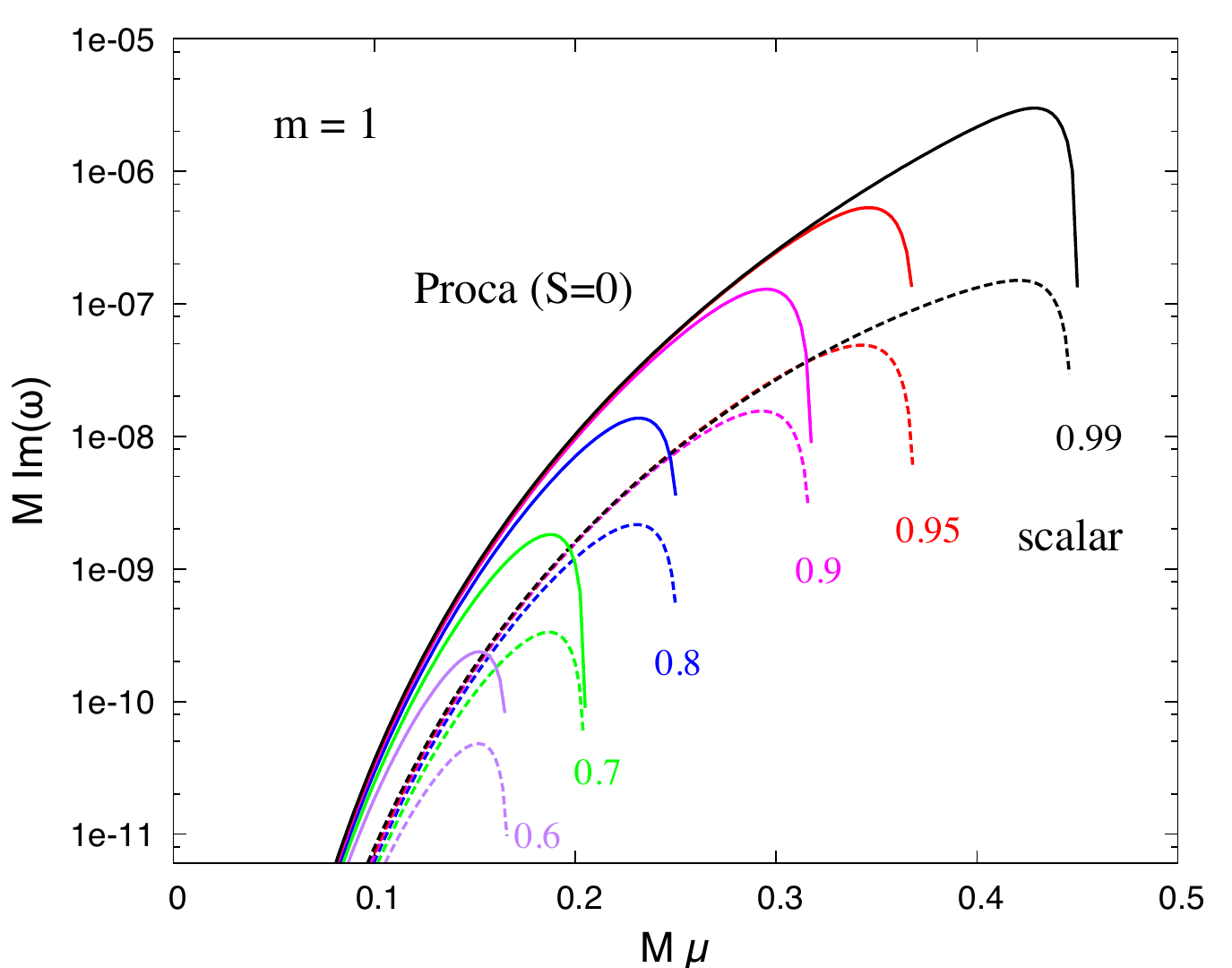}
\end{center}
\caption{
Comparing the growth rate of the `odd-parity' corotating dipole mode ($S=0$, $m = 1$) of the Proca field [solid], with the corotating dipole mode of the massive scalar field ($m = 1$, $\ell = 1$) [dashed]. The vertical axis shows the growth rate $\tau^{-1} = (GM/c^3) \text{Im}(\omega)$ on a logarithmic scale.
}
\label{fig:vectorscalar}
\end{figure}

Finally, Fig.~\ref{fig:vectorscalar} compares the growth rate of the $S=0$ (`odd-parity') $m=1$ fundamental mode of the Proca field, with the growth rate of the scalar field $l=m=1$ mode. The plot shows that the $S = 0$ mode of the Proca field displays qualitatively similar behaviour to the scalar field, but nevertheless has an enhanced growth rate, as anticipated due to the enhancement of superradiance with field spin. As noted earlier, the $S = -1$ mode grows much faster (by $\sim$ two orders of magnitude) than the $S = 0$ mode, because it has a greater binding energy (see Fig.~\ref{fig:re} and \ref{fig:wf}).

\section{Discussion\label{sec:discussion}}
We have calculated the instability rate for the Proca field on Kerr spacetime by solving the ordinary differential equations recently obtained by Frolov \emph{et al.}~\cite{Frolov:2018ezx}. We have presented data for the bound states of all three polarizations of the Proca field ($S = -1$, $0$, $+1$).

A key result of this work is that the `odd-parity' (axial) $S = 0$ modes can be recovered from the FKKS ansatz for the Proca field (N.B.~these modes were not identified in Ref.~\cite{Frolov:2018ezx} itself). A subtlety is that the eigenvalue $\nu$ diverges in the static limit $a \rightarrow 0$, although $a \nu$ has a well-defined limit. A confounding factor is the challenge of solving the angular equation via direct numerical integration for odd-parity eigenvalues with $a^2 \nu^2 > 1$, as $q_\theta$ passes through zero. However, the spectral method introduced in Sec.~\ref{sec:spectral} does not suffer from this issue. 

Three pieces of evidence suggest that we have correctly identified the axial ($S = 0$) modes. First, the radial equation (\ref{eq:R1}) reduces to Eq.~(8) in Ref.~\cite{Rosa:2011my} in the Schwarzschild limit. Second, the angular profile of $A^a$ also takes the correct form in this limit. Third, the growth rate of the bound states has a power-law scaling with the index anticipated by Eq.~(\ref{eq:omI}) for the $S=0$ mode.

In Fig.~\ref{fig:novertone} we showed that the $S = 0$ axial mode grows at a slower rate than several overtones of the $S = -1$ mode. This suggests that the $S = 0$ sector can be safely neglected in considerations of (hypothetical) astrophysical superradiant instabilities for vector bosons, and the phenomenology described in Refs.~\cite{Pani:2012vp, Baryakhtar:2017ngi, Cardoso:2018tly} is not altered in substance. 

Our numerical results for $S=-1$ and $S=+1$ are consistent with those presented by FKKS in Fig.~1 of Ref.~\cite{Frolov:2018ezx}, it would appear. Our numerical results are also consistent with those of Cardoso \emph{et al.} \cite{Cardoso:2018tly} for the dominant $S = -1$ mode, as shown in Fig.~\ref{fig:max}. However, our results for $S = 0$ and $+1$ are not consistent with those labelled $S=0$ and $+1$ in Ref.~\cite{Cardoso:2018tly}. We find that the $S = 0$, $m=1$ mode grows at a significantly slower rate than is implied by Fig.~6 of Ref.~\cite{Cardoso:2018tly}. Specifically, we find a maximum rate of $M \omega_I \sim 3 \times 10^{-6}$ for $a = 0.99M$, whereas Ref.~\cite{Cardoso:2018tly} find $M\omega_I \sim 1.4 \times 10^{-4}$ for $a \rightarrow M$. The latter rate is more consistent with the first excited state ($\hat{n}=1$) of the $S = -1$ mode, shown in Fig.~\ref{fig:novertone} (which we find takes the value of $M\omega_I \sim 7.6 \times 10^{-5}$ for $a/M = 0.99$), suggesting that the modes shown in Fig.~6 of Ref.~\cite{Cardoso:2018tly} are all $S = -1$ modes.

Our investigation into the angular spectrum is not comprehensive, and is centred around special cases for $\omega = \pm \mu$. Based on our analysis, we cannot yet rule out the possibility that other branches of solution exist, perhaps corresponding to unexpected polarizations. Further work is needed, for example, to find series expansions of the eigenvalues $\nu$ in $a\omega$ and $a\mu$.

Some further attention could be given to the stationary modes ($\omega_I = 0$) that exist at the superradiant cut-off $\omega_R = m\Omega_H$ \cite{Hod:2012px}. Such modes are closely linked to the family of `hairy' black holes with Proca field hair identified by Herdeiro, Radu and Runarsson \cite{Herdeiro:2014goa,Herdeiro:2015waa,Herdeiro:2016tmi}.

To conclude, we have found that the separation of variables achieved by Frolov \emph{et al.}~\cite{Frolov:2018ezx, Krtous:2018bvk, Frolov:2017kze} makes it rather straightforward to study the superradiant instability that afflicts the Proca field on the Kerr spacetime. All three polarizations of the Proca field can be found via ansatz (\ref{eq:ansatz}). We anticipate future works on vector bosons interacting with (higher-dimensional) Kerr-(A)dS-NUT fields will exploit new-found separability properties to great advantage. 

\begin{acknowledgments}
With thanks to Asimina Arvanitaki, Masha Baryakhtar, William East and Robert Lasenby for organising the meeting  ``{\it Searching for New Particles with Black Hole Superradiance}'' held at Perimeter Institute for Theoretical Physics on 9th--11th May 2018. Research at Perimeter Institute is supported by the Government of Canada through Industry Canada and by the Province of Ontario through the Ministry of Economic Development \& Innovation. With additional thanks to David Kubiz\v{n}\'ak and Jo\~ao Rosa. I acknowledge financial support from the European Union's Horizon 2020 research and innovation programme under the H2020-MSCA-RISE-2017 Grant No.~FunFiCO-777740, and from the Science and Technology Facilities Council (STFC) under Grant No.~ST/L000520/1.
\end{acknowledgments}

\appendix

\section{Killing tensors in Carter's tetrad\label{appendix:a}}
Here we list some explicit expressions for the Killing quantities on Kerr spacetime \cite{Frolov:2017kze}. 
Carter's canonical tetrad (closely related to the Darboux basis) is
\beq
\omega_{(0)}^a = \sqrt{\frac{1}{\Sigma \Delta_r}} \left(\partial_\psi^a + r^2 \partial_\tau^a \right), 
\quad \omega_{(1)}^a = \sqrt{\frac{\Delta_r}{\Sigma}} \partial_r^a, 
\quad \omega_{(2)}^a = - \sqrt{\frac{\Delta_y}{\Sigma}} \partial_y^a , 
\quad \omega_{(3)}^a = \sqrt{\frac{1}{\Delta_y \Sigma}} \left(-\partial_\psi^a + y^2 \partial_\tau^a \right).
\eeq
such that $g_{ab} \omega_{(\alpha)}^a  \omega_{(\beta)}^b  = \eta_{\alpha \beta}$. In this basis,
\begin{subequations}
\begin{eqnarray}
g^{ab} &=& -\omega_{(0)}^a \omega_{(0)}^b + \omega_{(1)}^a \omega_{(1)}^b + \omega_{(2)}^a \omega_{(2)}^b + \omega_{(3)}^a \omega_{(3)}^b , \\
h^{ab} &=& \phantom{+} 2 r \omega_{(0)}^{[a} \omega_{(1)}^{b]} - 2 y \omega_{(2)}^{[a} \omega_{(3)}^{b]} , \\
f^{ab} &=& -2 y \omega_{(0)}^{[a} \omega_{(1)}^{b]} - 2 r \omega_{(2)}^{[a} \omega_{(3)}^{b]} . \\ 
Q^{ab} &=& -r^2 \left(-\omega_{(0)}^a \omega_{(0)}^b +  \omega_{(1)}^a \omega_{(1)}^b \right) + y^2 \left(\omega_{(2)}^a \omega_{(2)}^b + \omega_{(3)}^a \omega_{(3)}^b \right) , \\
K^{ab} &=& -y^2  \left(-\omega_{(0)}^a \omega_{(0)}^b +  \omega_{(1)}^a \omega_{(1)}^b \right) + r^2 \left(\omega_{(2)}^a \omega_{(2)}^b + \omega_{(3)}^a \omega_{(3)}^b \right)  .
\end{eqnarray}
\end{subequations}

\bibliographystyle{apsrev4-1}
\bibliography{proca_kerr_refs}

\end{document}